# Climatological variability of a thunderstorm environment dataset in tropical and temperate regions

Andrew Dowdy*[1,2], Andrew Brown[1,2], Todd P. Lane[1,2], Mateusz Taszarek[3]

[1]School of Geography, Earth and Atmospheric Sciences, The University of Melbourne, Melbourne, Australia

[2]ARC Centre of Excellence for 21st Century Weather, The University of Melbourne, Melbourne, Australia

[3]Department of Meteorology and Climatology, Adam Mickiewicz University, Poznań, Poland

*andrew.dowdy@unimelb.edu.au; orcid.org/0000-0003-0720-4471

## Abstract

Spatiotemporal variations in thunderstorm occurrence frequency are considered here using an environmental dataset derived from ERA5 reanalysis data. Interannual variability in the thunderstorm environments is examined for the period 1979–2023, with the standard deviation and coefficient of variation showing considerable spatial differences through the world. Atmospheric and oceanic modes of climate variability account for some of this interannual variability, particularly for the El Niño-Southern Oscillation through tropical and maritime regions, as well as to a lesser degree for the Indian Ocean Dipole, Arctic Oscillation and Antarctic Oscillation. Long-term trends can also contribute to interannual variability, with results showing increases are more common than decreases in the thunderstorm environments through the study region over the period 1979–2023. However, considerable uncertainties in those trends are noted as is also suggested from some additional analysis of global climate models, indicating that although more favorable thunderstorm environments might occur in a warming world, the estimated change over the period 1979–2023 is relatively small compared to the standard deviation in most locations. The study findings are intended to be complementary to other studies and contribute as part of a broader range of information available on thunderstorms and climate variability.

# 1 Introduction

Many studies have examined thunderstorms and the hazards they can cause such as lightning, hail, heavy rain, damaging wind gusts, turbulence and tornadoes (Droegemeier and Wilhelmson 1987; Doswell III 2003; Lane et al. 2012; Virts et al. 2013; Tippett et al. 2015; Allen 2018; Prein and Holland 2018; Lavigne et al. 2019). Those phenomena can also contribute to other hazards such as flash floods (Maddox et al. 1980; Davis 2001; Wasko et al. 2024), as well as lightning accompanied by little or no rainfall on the ground known as 'dry lightning' which is an important ignition source for wildfires (Rorig et al. 2007; Dowdy and Mills 2012; Kalashnikov et al. 2022). Lightning from thunderstorms also plays a key role in producing various atmospheric components that influence greenhouse gases and chemistry, including ozone and oxides of nitrogen (Luhar et al. 2021). It is therefore important for many reasons to understand the climatological variability of thunderstorm occurrence.

There are considerable challenges in examining the thunderstorm climatology through the world, because homogenous long-term observations of thunderstorms are not widely available globally, as well as because the resolution of climate models currently available is not able to accurately simulate fine-scale processes that cause thunderstorms (e.g., processes associated with convective initiation and microphysics such as cloud condensation, interactions of different hydrometeor species and turbulent mixing) (Droegemeier and Wilhelmson 1987; Hoogewind et al. 2017; Gutowski et al. 2020). As such, diagnostic methods are sometimes used for climate analysis purposes based on environmental proxies from coarser grid datasets that indicate favorable ingredients for convection and thunderstorm development, noting a wide range of diagnostics methods presented in many previous studies (Romps et al. 2014; Kent et al. 2015; Finney et al. 2018; Prein and Holland 2018; Lepore et al. 2021).

Some examples of commonly-used environmental diagnostics of thunderstorm occurrence include convective available potential energy (CAPE) and vertical wind shear from the surface to 6 km above ground level (WS06) (Brooks et al. 2003; Craven and Brooks 2004; Gensini and Ashley 2011; Allen and Karoly 2014; Westermayer et al. 2017; Dowdy 2020; Lepore et al. 2021; Taszarek et al. 2021a; Battaglioli et al. 2023). Higher values of CAPE are typically associated with warm and moist air at lower heights of the atmosphere and relatively cold air above, providing large vertical gradients of temperature (lapse rate). CAPE is associated with the potential maximum updraft speed in thunderstorms based on the theoretical relationship for potential energy represented by CAPE being converted to kinetic energy (e.g., Williams and Renno (1993)). Similarly, higher values of vertical wind shear (WS06) can be favorable for thunderstorm development, maintenance and intensification processes, including to help tilt the storm system and separate updrafts from falling precipitation, which lead to better storm organization, longer duration and overall higher potential for producing convective hazards (Weisman and Klemp 1982; Markowski and Richardson 2011; Mulholland et al. 2024). Wind shear may also contribute to thunderstorm occurrence in some cases reported to have low values of CAPE (Allen and Karoly 2014; Sherburn and Parker 2014; King et al. 2017).

A combination of CAPE and WS06 in a single diagnostic proxy has been shown in the past to be skilful in determining likelihood of convective storm occurrence worldwide (Brooks et al. 2003; Craven and Brooks 2004; Allen and Karoly 2014; Taszarek et al. 2020; dos Santos et al. 2023). This study uses the Broadscale Thunderstorm Environment (BTE) dataset that is based on a diagnostic method for indicating environmental conditions conducive to thunderstorm occurrence as described in Dowdy and Brown (2025a), with further details on development also documented in Dowdy and Brown (2023; 2025b). The diagnostic is derived from CAPE and WS06 calculated from the ERA5 reanalysis (Hersbach et al. 2020), with lower limits applied for CAPE and WS06, as described in Section 2.

The aim of this study is to examine aspects of the interannual variability in thunderstorm occurrence as represented by this environmental diagnostic method. The influence of modes of climate variability is also examined, including for the El Niño-Southern Oscillation (ENSO), Indian Ocean

Dipole (IOD), Arctic Oscillation (AO; also known as the Northern Annual Mode, NAM) and the Antarctic Oscillation (AAO; also known as the Southern Annular Mode, SAM). Additionally, the potential for long-term trends is examined, noting theoretical expectations for potential increases in thunderstorm activity in a warmer world including associated with Clausius Clapeyron considerations for increased water vapour concentrations (Romps et al. 2014; Hoogewind et al. 2017; Allen 2018; Dowdy 2020; Lepore et al. 2021; Cheng et al. 2022; Allan et al. 2023). Data and methods are detailed in Section 2, results presented in Section 3 and discussed in Section 4, with some additional information in the Appendix.

## 2 Data and methods

### 2.1 Thunderstorm environments based on ERA5 reanalysis

The BTE dataset is available as described in Dowdy and Brown (2025a) based on the ERA5 reanalysis produced by the European Centre for Medium-Range Weather Forecasts (ECMWF) (Hersbach et al. 2020). The BTE dataset covers the period from 1979 to 2023 at 6-hourly time steps, representing instantaneous values for 0000, 0600, 1200 and 1800 UTC. It is on a 0.25-degree grid that is global in longitude, and from 69.25°N to 69.25°S in latitude, defining the study region used here. The dataset is based on a diagnostic quantity as shown in Equation 1, similar to an original idea of the Significant Severe Parameter (SSP) from Craven and Brooks (2004), but also applying lower limits of CAPE $\geq$ 10 J.kg$^{-1}$ and WS06 $\geq$ 10 m.s$^{-1}$ as described in Dowdy and Brown (2023). Including these lower limits helps allow for thunderstorms that may occasionally occur in low-CAPE or low-shear environments (Sherburn and Parker 2014; King et al. 2017; Westermayer et al. 2017; Miller and Mote 2018; Prein and Holland 2018), while making little difference to the general spatial features of the diagnostic (e.g., comparing results in Dowdy and Brown (2023) for their Figures 10 and 14). Dowdy and Brown (2023) also showed that the diagnostic is robust to some changes in the limiting values used, such as finding very similar results if those limits are halved which would result in a higher fraction of values being raised to the threshold (e.g., their Appendix Fig. A7).

BTE = CAPE * WS06  (Equation 1)

where CAPE $\geq$ 10 J.kg$^{-1}$ and WS06 $\geq$ 10 m.s$^{-1}$ (i.e., all values lower than those limits are set equal to 10).

For the BTE dataset as used in this study, CAPE was calculated based on the most unstable air parcel using the wrf-python software package under irreversible pseudo-adiabatic ascent with a virtual temperature correction applied (Ladwig 2017). Thunderstorms are estimated to have occurred when the diagnostic value (i.e., *BTE* as shown in Equation 2) exceeds a threshold. The threshold at each grid location is the value of *BTE* that is exceeded as frequently as the occurrence of the observations-based thunderstorm data, with World Wide Lightning Location Network (WWLLN) lightning data (Virts et al. 2013) from 2012 to 2023 used as a proxy for thunderstorm observations. This means that the threshold value at each location is set such that the number of thunderstorm environments indicated (i.e., by *BTE* exceeding the threshold value) is equal to the number of observed thunderstorms (based on lightning observations). Consequently, a feature of this type of method using spatially-varying diagnostic thresholds is that it produces results consistent with observations in terms of the average occurrence frequency, such as has been demonstrated for historical thunderstorms in Australia (Dowdy 2020; Pepler et al. 2021) and other regions through the world (Dowdy and Brown 2023).

Further details on the method development were documented in Dowdy and Brown (2023), similar to the previous version of this broadscale thunderstorm environment dataset (Dowdy 2020), with the observed thunderstorms considered for the purposes of the BTE dataset based on 2 or more lightning strokes being recorded within ± 3 grid cells of a given location during a 6-hourly time period. As stated in Dowdy and Brown (2023), that method of aggregating lightning observations within 0.75 degrees around each grid cell and in 6-hourly time periods was designed for the purpose of representing broadscale environmental conditions including as this helps facilitate application of the method to a range of data sets (including coarse-scale climate model data: see Section 2.4) and allowing for variations between different convective systems such as in their speed of movement over a region. They further stated that this approach was designed to be complementary to other methods suitable for finer-scale model data and diagnostics including those based on microphysical processes (Lopez 2016).

The probability of detection (POD) based on this diagnostic approach is shown in Appendix Fig. 8, consistent with previous studies that documented the development of this method used here (Dowdy 2020; Dowdy and Brown 2023, 2025b), noting higher values generally for land and tropical ocean regions with relatively low values in higher latitudes such as towards Antarctica as well as eastern ocean basins in the midlatitudes. Results are interpreted accordingly through this study, including considering the regions noted here with relatively low POD. Reasons for the variation in POD values include CAPE and shear being just two of the many conditions that can sometimes be associated with thunderstorm occurrence. For example, CAPE is typically a necessary, but not sufficient, condition for thunderstorm occurrence, in addition to other factors such as lifting mechanisms that can help initiate the release of CAPE, as well as noting a range of other physical processes that may also contribute (e.g., aerosols for cloud condensation nuclei and turbulence mixing (Singh and O'Gorman 2013; Peters et al. 2023)).

As noted above, CAPE data used in this study was calculated here using code from the wrf-python software package (Ladwig 2017). This is beneficial in allowing that same code to also be applied to other data sets for consistency, including for application to global climate models (GCMs) as discussed in Section 2.4. CAPE calculated in this way appears to show more similarities to previous analysis based on radiosonde observations, in that it doesn't show the general tendency for a downward trend that occurs for CAPE obtained directly from the ERA5 reanalysis dataset particularly in some tropical ocean regions (Taszarek et al. 2021a,b). Examples of the different trends are shown in Appendix Fig. 9 for CAPE calculated ourselves as used in this study, as compared to Appendix Fig. 10 for CAPE as obtained directly from ERA5. While the cause for these differences in CAPE is not unknown, the CAPE calculated ourselves is preferred for the purposes of this study for reasons as noted here, i.e., allowing consistent application of the wrf-python software package (Ladwig 2017) to different datasets, as well as because it represents convective environments well compared to observations and doesn't show the tendency for downward trends that oppose radiosonde-based results (Taszarek et al. 2021a,b; Pilguj et al. 2022).

Orography data (land-sea mask) from ERA5 reanalysis (Hersbach et al. 2020) is used to classify grid cells as land or ocean through the study region). This is also used for all coastline mapping in figures throughout the study.

## 2.2 Atmospheric and oceanic modes of climate variability

The relationship between the thunderstorm environments and four different modes of atmospheric and oceanic variability is examined in this study. The modes of variability considered here are as follows: ENSO based on the NINO3.4 index (Rasmusson and Carpenter 1982); the Indian Ocean Dipole (IOD) based on the Dipole Mode Index (DMI) (Saji et al. 1999); the Arctic Oscillation (AO) (Thompson and Wallace 1998) and the Antarctic Oscillation (AAO) (Gong and Wang 1999). Index data for NINO3.4, DMI, AO and AAO were obtained from National Oceanic and Atmospheric Administration (NOAA) from the website http://www.cpc.ncep.noaa.gov/data/indices/sstoi.indices (accessed December 2024).

These climate modes represent key influences on variability from seasonal to interannual time scales for this study region throughout tropical and midlatitude regions of the world, showing relationships to environmental conditions associated with thunderstorm occurrence in various regions through the world (Yoshida et al. 2007; Kulkarni et al. 2013; Allen and Karoly 2014; Pinto Jr 2015; Dowdy 2016). For example, ENSO and IOD are ocean-atmosphere coupled modes of variability, associated with spatial variations in sea surface temperature that interact with atmospheric circulation (Rasmusson and Carpenter 1982; Saji et al. 1999), with AO and AAO characterised by north-south shifts in atmospheric mass between the polar regions and the middle latitudes (Thompson and Wallace 1998; Gong and Wang 1999).

The Pearson's correlation coefficient, $r$, is used to estimate relationships between the occurrence of thunderstorm environments and the indices representing the modes of climate variability (i.e., NINO3.4, DMI, AO and AAO). The relationships are examined for individual seasons: December to February (DJF), March to May (MAM), June to August (JJA) and September to November (SON). This is done using data from December 1979 to November 2023 to create seasonal values from one year to the next (i.e., 43 years of seasonal values), including for the number of thunderstorm environments identified at each location as well as for the index values representing each of the modes of climate variability. The correlations are calculated individually for each season at each individual grid-cell location, with $|r| > 0.25$ representing a confidence level of 90% based on these data (i.e., two-tailed probability $p < 0.1$ of no relationship).

### 2.3 Statistical analysis of variability

To examine the potential for long-term changes in the climatology of thunderstorm environments, the method used here is based on comparing the mean value for the first half of a time period with the mean value for the second half of that time period. Statistical significance in these results is indicated using a bootstrap method that ranks the magnitude of a change (i.e., the difference in mean values from the first half to the second half of the time period) with a 1000-member randomized sample of years used to calculate the mean values.

Results are presented for a confidence level above 90% based on this two-tailed nonparametric approach, noting that this 90% threshold is a somewhat arbitrary choice. The statistical significance test (i.e., using bootstrapping as noted above) is applied individually to data for each grid cell to test for local significance, rather than applying the statistical testing to data combined from multiple locations to test for field significance (Livezey and Chen 1983). Additionally, it is acknowledged that 10% of cases on average will be indicated as being significant due to random chance alone at the 90% confidence level used throughout this study.

### 2.4 Thunderstorm environment data based on global climate models

A version of the BTE dataset is also available based on 12 global climate models (GCMs) for historical (1979 to 2005) and future (2089 to 2100) time periods. The models are from the CMIP5 ensemble (Taylor et al. 2012) using a high emissions pathway (RCP8.5) which helps enhance the signal to noise ratio for quantifying climate change influences, similar to approaches in previous studies (Finney et al. 2018). Those are the only models the BTE dataset is currently available for, with those model data being used only due to having the 6-hourly data readily available during the development process of the dataset as described in Dowdy and Brown (2025a,b). Potential to build on this in future research is also noted, e.g., using other models and emissions pathways such as based on CMIP6 GCMs (O-Neill et al. 2016).

The average global surface temperature for each GCM and time period is listed in Table 1 (calculated from data available here https://github.com/traupach/warming_levels) and is used for each GCM to calculate the changes per degree of warming in the number of thunderstorms indicated by the environmental diagnostic. CAPE and WS06 were produced using the same method as detailed in Section 2.1 for the dataset version based on ERA5 reanalysis, before being bilinearly interpolated to the same 0.25-degree grid in latitude and longitude. The global climate model data for CAPE and WS06 were calibrated using quantile-quantile matching for the historical period 1979 to 2005, using the data based on ERA5 reanalysis as the reference. The quantile matching was done using Python software code based on an existing approach (Seabold and Perktold 2010) (see Data Availability section for further details).

**Table 1** The 12 climate models used in this study and their global average surface temperatures (to one decimal place). This is presented for the historical period from 1979 to 2005, as well as for the future period from 2081 to 2100 under a high emissions scenario

| Model name | 1979:2005 | 2081:2100 |
|---|---|---|
| ACCESS1-0 | 15.1°C | 19.2°C |
| ACCESS1-3 | 15.4°C | 19.4°C |
| BNU-ESM | 13.9°C | 18.3°C |
| CNRM-CM5 | 13.9°C | 17.5°C |
| GFDL-CM3 | 13.8°C | 18.5°C |
| GFDL-ESM2G | 13.8°C | 16.7°C |
| GFDL-ESM2M | 14.3°C | 17.0°C |
| IPSL-CM5A-LR | 13.0°C | 17.6°C |
| IPSL-CM5A-MR | 13.8°C | 18.3°C |
| MIROC5 | 14.9°C | 18.4°C |
| MRI-CGCM3 | 14.0°C | 17.2°C |
| bcc-csm1-1 | 14.6°C | 18.1°C |

## 3  Results

### 3.1  Interannual variability

The standard deviation in the annual number of thunderstorm environments is shown in Fig. 1, based on the BTE dataset at 6-hourly time steps during the period from 1979 to 2023. The regions with relatively large variability tend to be in the tropics, rather than the higher latitudes in each hemisphere where the variability tends to be relatively small. In the midlatitudes, relatively high variability tends to occur for land regions and on the western side of ocean basins, with relatively low variability on the eastern side of midlatitude ocean basins. This is the case for the North Pacific, South Pacific, North Atlantic, South Atlantic and Indian ocean basins in this study region.

Although the regions with relatively high variability (from Fig. 1a) also tend to have relatively large mean values, as shown in Fig. 1b, there is some variation in that relationship as shown by the coefficient of variation (Fig. 1c). The coefficient of variation is calculated as the ratio of the standard deviation divided by the mean, calculated individually at each location, as a measure to indicate the magnitude of interannual variability as a fraction of the mean.

The coefficient of variation tends to be larger in regions with relatively low mean values. Those regions include many of the higher latitude locations, as well as some midlatitude locations on the eastern side of ocean basins. Regions with relatively small values for the coefficient of variation occur in the tropics for some land regions and in the Maritime Continent, as well as in some midlatitude regions over land and on the western side of ocean basins.

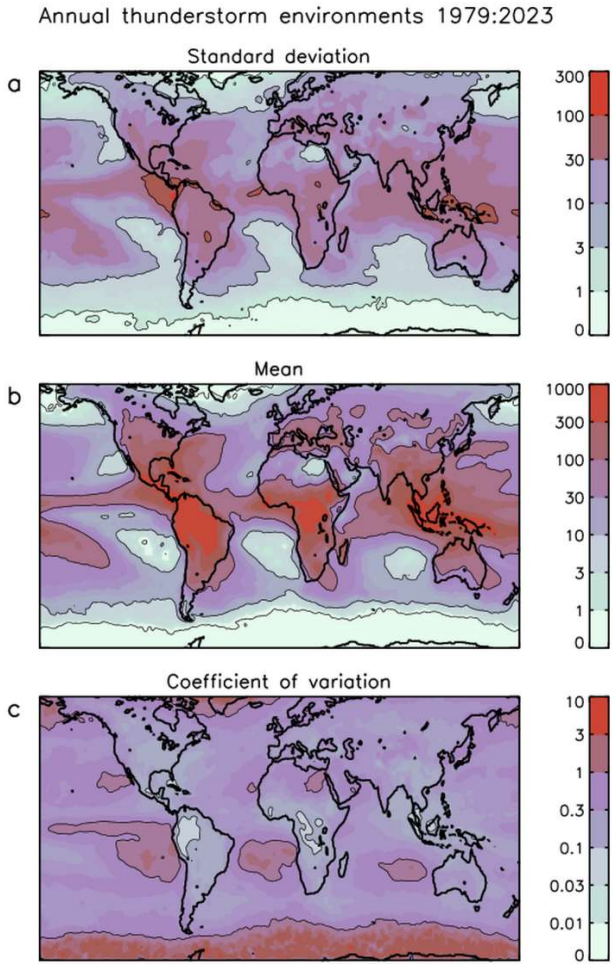

**Fig 1** Variability in the annual number of 6-hourly thunderstorm environments per grid cell during the period from 1979 to 2023. Results are shown for the standard deviation (**a**), mean (**b**), and coefficient of variation (**c**). Contour lines are shown for values of 0.1, 1, 10 and 100 counts, with smoothing applied to the results shown in each panel (using a boxcar moving average of ±11 grid cells in latitude and longitude)

### 3.2  Influence of modes of climate variability

There are many sources of natural variability, including large-scale atmospheric and oceanic modes that can cause significant fluctuations in weather and climate. As described in Section 2.2, four different modes of climate variability are examined here in relation to the occurrence frequency of favorable thunderstorms environments: ENSO, IOD, AO and AAO.

Figure 2 presents correlations between the NINO3.4 index (as a measure of ENSO) and thunderstorm environment occurrence frequency based on the diagnostic method applied to ERA5 reanalysis data. The Pearson correlation coefficient, $r$, is calculated individually for each location and for four seasons: December, January and February (DJF), March, April and May (MAM), June, July and August (JJA) and September, October and November (SON). The results show that ENSO has a significant influence on thunderstorm environments in many tropical regions of the world, as well as in some extratropical regions (Fig. 2). About 25% of locations through the study region have significant correlations in each season (i.e., based on the 90% statistical confidence level, 2-tailed, used throughout this study as detailed in Section 2).

Table 2 quantifies these results individually for the tropics and extratropics, as well as for land and ocean regions. This shows that significant relationships for ENSO occur more extensively in the tropics (ranging from 32-52% of the study region) than the extratropics (ranging from 7-23% of the study region). Higher values are also seen for ocean regions (ranging from 14-52% of the study region) than land regions (ranging from 7-36% region), while noting the diagnostic tends to have somewhat lower skill in ocean regions than land regions in general (e.g., see Appendix Fig. 8).

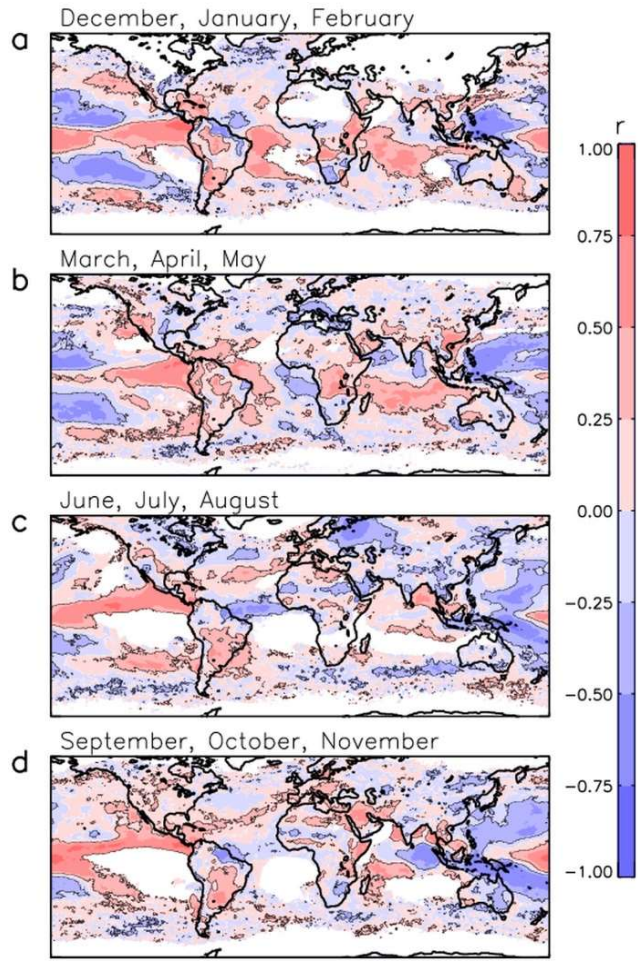

**Fig 2** Correlations between seasonal values of ENSO (NINO3.4 index) and the seasonal number of thunderstorm environments per grid cell, using data from December 1979 to November 2023. The correlations are calculated individually at each grid location and for each season: December, January and February (**a**); March, April May (**b**); June, July, August (**c**); September, October and November (**d**). For each season and location, results are shown only when at least 10 years have non-zero counts of the thunderstorm environments, with white regions representing locations that don't meet that criterion. The thin black contour line encloses locations where the magnitude of the correlation, *r*, is statistically significant (90% confidence level)

The relationships between the thunderstorm environments and the other three modes of climate variability are shown in Figures 3-5, for the IOD, AO and AAO, respectively. The relationships for these other modes of variability are not as strong as the case for ENSO (from Fig. 2). For the example of the IOD (Fig. 3), the relationships are strongest during the SON season, while noting this occurs in approximately the same locations seen for ENSO (from Fig. 2), with somewhat fewer locations having significant relationships during the other seasons as compared to SON. The similarity between the IOD and ENSO results during the SON season is to be expected to some degree, given the relationship between ENSO and IOD during SON as shown in Table 3 for seasonal correlations between the climate modes. The relationship between ENSO and IOD could relate in part to feedback processes between the tropical Indian Ocean and Pacific Ocean linked through convection in the Maritime Continent region. The AO and AAO results (Figures 4 and 5) show some regions of significant correlations, including at higher latitudes. Some of those high latitude locations with significant AO and AAO correlations are different to the locations where ENSO correlations were significant (from Fig. 2), suggesting additional information can be provided by those polar annular modes of variability.

Table 2  Relationships between the number of thunderstorm environments per grid cell that occurs in a season and metrics for each of the climate modes (ENSO, IOD, AO and AAO). Values are shown for the percentage of the study region that has statistically significant correlations as shown in Figures 2-5. This is presented individually for each season (DJF, MAM, JJA and SON) based on data during the period from December 1979 to November 2023, including for the tropical and extratropical latitudes and for different orographic locations (i.e., land or ocean regions)

| Mode | Latitudes | Orography | DJF | MAM | JJA | SON |
|---|---|---|---|---|---|---|
| ENSO | Tropics | Land | 34% | 33% | 32% | 36% |
|  |  | Ocean | 51% | 52% | 37% | 44% |
|  | Extratropics | Land | 7% | 13% | 23% | 18% |
|  |  | Ocean | 16% | 15% | 18% | 14% |
| IOD | Tropics | Land | 20% | 25% | 24% | 43% |
|  |  | Ocean | 20% | 11% | 19% | 37% |
|  | Extratropics | Land | 4% | 13% | 19% | 22% |
|  |  | Ocean | 12% | 9% | 11% | 13% |
| AO | Tropics | Land | 9% | 18% | 10% | 9% |
|  |  | Ocean | 4% | 15% | 12% | 6% |
|  | Extratropics | Land | 6% | 10% | 14% | 6% |
|  |  | Ocean | 14% | 10% | 13% | 5% |
| AAO | Tropics | Land | 19% | 24% | 5% | 19% |
|  |  | Ocean | 20% | 11% | 8% | 17% |
|  | Extratropics | Land | 8% | 13% | 7% | 13% |
|  |  | Ocean | 15% | 14% | 14% | 13% |

There are only two cases where other modes of variability have values higher than ENSO for the fraction of the study area that has significant correlations, as shown in Table 2. The first case is for the AAO in extratropical land regions during the DJF season with a value of 8%. However, that is only 1% larger than the value of 7% for ENSO in that case, with these values not likely to be meaningful given that on average about 10% of the region could be indicated as significant due to random chance alone (as noted in Section 2 given the 90% confidence level used). The second case is also for extratropical land regions, but for the IOD during the SON season with a value of 22%, which is 4% larger than the value of 18% for ENSO. However, as noted above, ENSO is closely related to IOD during spring (with a correlation of $r = 0.63$ from Table 3), with the regions with significant correlations during SON season for IOD (Fig. 3d) occurring in broadly similar locations to those for ENSO (Fig. 2d).

The AAO has a relatively high value for MAM in land regions, which appears primarily associated with a negative correlation in large parts of South America (from Fig. 5b) which extends to the southeast into a larger band of negative correlations around the extratropics. One potential physical explanation for that the extratropical band with primarily negative correlations might relate to latitudinal shifts in storm tracks, as higher AAO values are typically associated with a stronger and tighter polar vortex including a more poleward position of the midlatitude storm tracks with fewer storms in those midlatitude regions for higher AAO.

Based on the results shown in Table 2, as well as Figures 2-5, ENSO clearly has a more extensive influence on thunderstorm environments through the study region overall as compared to the three other modes of variability examined here. This is broadly similar to previous studies that show ENSO is a dominant mode of natural climate variability throughout the world, including for thunderstorm activity such as shown in previous studies using different data sets and methods complementary to those used here (Allen and Karoly 2014; Pinto 2015; Dowdy 2016; Tippett and Lepore 2021; Malloy et al. 2023). The results presented here also highlight that the ENSO influence is particularly strong and widespread in tropical regions as well as in ocean regions in general, with somewhat less differences between ENSO and other modes of variability for extratropical and land regions. The additional information provided by the polar annular modes (i.e., AO and AAO) in high latitudes is also notable based on these results.

For some additional examples of variations relating to ENSO, Appendix Figures 11-13 show time series and scatter plots are shown for two individual regions, selected on the western (Torres Strait) and eastern (Galapagos Islands) sides of the Pacific Ocean. The example for the Torres Strait region (around the northeast of Australia as shown in Appendix Fig. 11a) is presented for the SON season, with that season having the strongest correlation as shown in Table 3. The correlation is negative for that Torres Strait region as shown in Appendix Fig. 12, consistent with the broad region of negative correlations near northeast Australia for SON as also shown here in Fig. 3d. The second example for the Galapagos Island region (in the tropics to the west of South America as shown in Appendix Fig. 11b) is presented for the DJF season, as a contrasting case to the previous example for the western Pacific in SON. The correlation is positive for that Galapagos Island region as shown in Appendix Fig. 13, consistent with the broad region of positive correlations for the tropical eastern Pacific Ocean region in DJF shown here in Fig. 3a. The relationships are shown for both of those examples using time series as well as scatter plots for the thunderstorm environments and NINO3.4 values (to represent ENSO conditions), just as examples to complement the results presented here in Fig. 3.

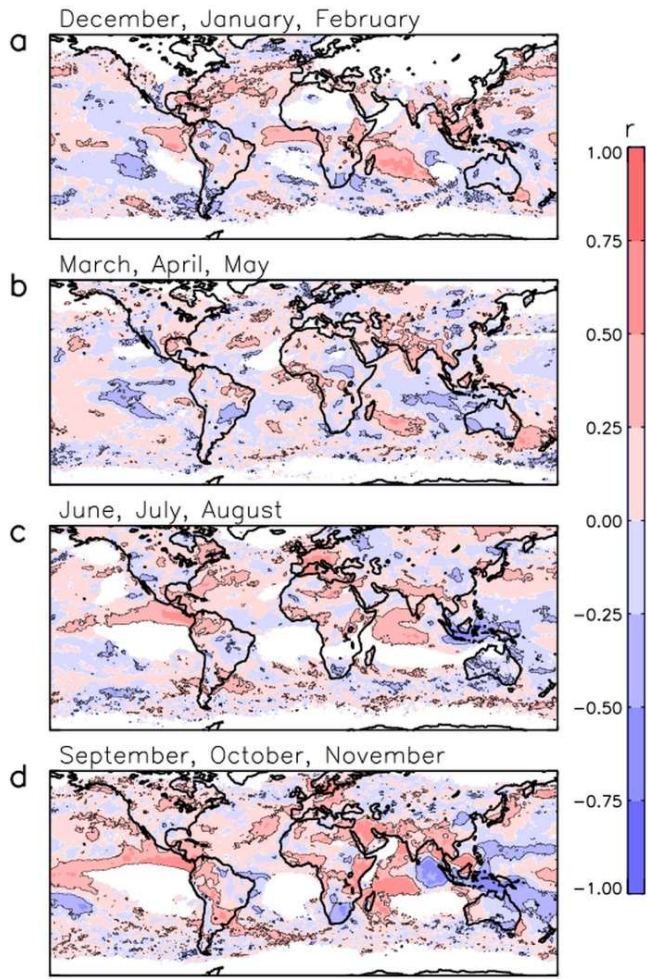

**Fig 3** As for Fig. 2, but for correlations between seasonal values of IOD (DMI index) and the seasonal number of thunderstorm environments. The correlations are calculated individually at each location and for each season: December, January and February (**a**); March, April May (**b**); June, July, August (**c**); September, October and November (**d**)

Table 3  The correlation between different large-scale modes of variability. This is shown for different pairs of indices representing the modes of variability (for ENSO, IOD, AO and AAO) and for each of the four seasons (DJF, MAM, JJA and SON), based on the period from December 1979 to November 2023

|  | DJF | MAM | JJA | SON |
|---|---|---|---|---|
| ENSO and IOD | 0.27 | -0.03 | 0.39 | 0.63 |
| ENSO and AO | -0.12 | -0.03 | -0.25 | 0.00 |
| ENSO and AAO | -0.44 | -0.06 | -0.02 | -0.28 |
| IOD and AO | -0.09 | 0.14 | -0.16 | 0.16 |
| IOD and AAO | 0.30 | 0.27 | 0.10 | -0.16 |
| AO and AAO | 0.17 | -0.11 | 0.06 | 0.25 |

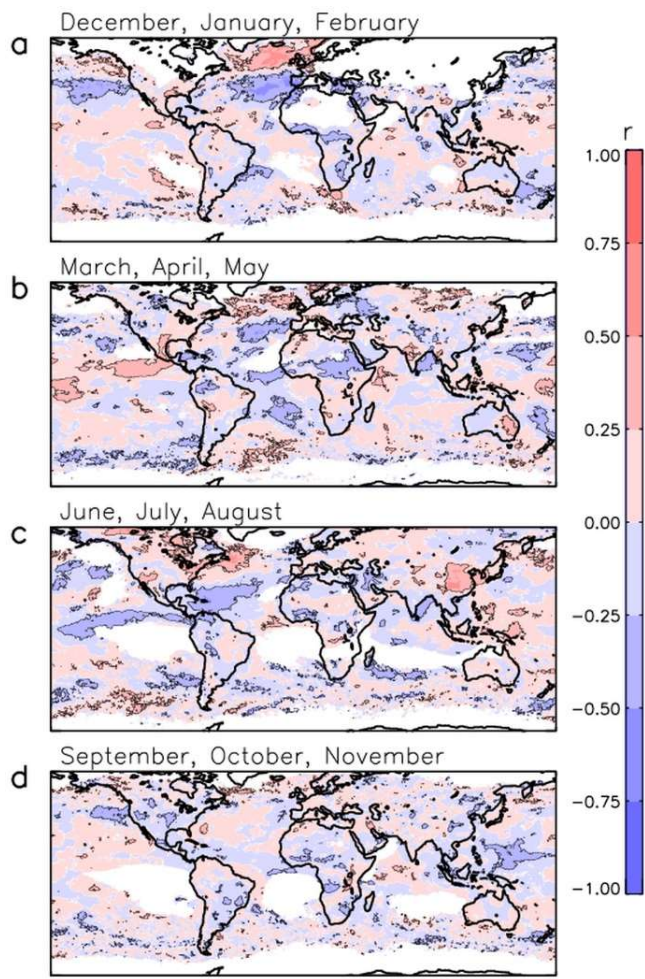

**Fig 4** As for Fig. 2, but for correlations between seasonal values of AO and the seasonal number of thunderstorm environments. The correlations are calculated individually at each location and for each season: December, January and February (**a**); March, April May (**b**); June, July, August (**c**); September, October and November (**d**)

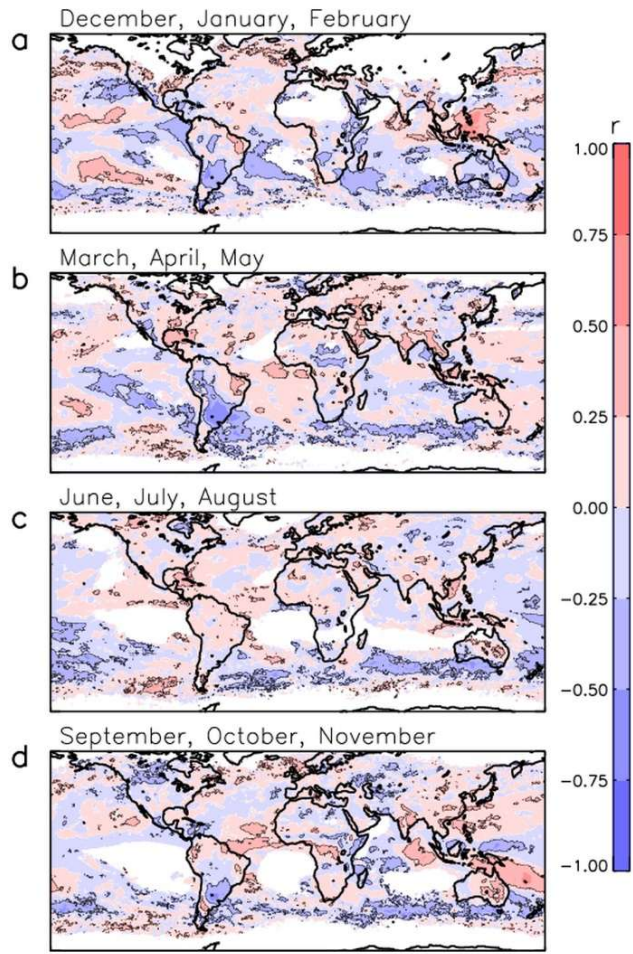

**Fig 5** As for Fig. 2, but for correlations between seasonal values of AAO and the seasonal number of thunderstorm environments. The correlations are calculated individually at each location and for each season: December, January and February (**a**); March, April May (**b**); June, July, August (**c**); September, October and November (**d**)

### 3.3 Long-term historical trend analysis

Figure 6 presents the average annual number of thunderstorms based on the environmental diagnostic during the first and second halves of the period from 1979 to 2023 (i.e., from 1979 to 2000 and from 2002 to 2023, with the central year of 2001 not used given the odd number of years). For the change from the first time period to the second time period, the results show that 15% of locations have significant increases and 8% have significant decreases through the study region (based on the 90% statistical confidence level).

The regions of increase are mostly located in the Northern Hemisphere, as well as some localized midlatitude regions on the west side of the ocean basins in both hemispheres (such as near southeast Australia). The regions of decrease include some land regions in each of the continents as shown (not including Antarctica), as well as a relatively large area of the tropics in the central-eastern Pacific Ocean where decreases are also indicated. However, given the degree of interannual variability (from Fig. 1) and other uncertainties such as based on statistical confidence testing (as discussed in Section 2), caution is recommended when interpreting these trend results particularly for small geographic regions.

Considering the study region overall, the results presented in Fig. 6 suggest some potential for an increase being more likely than a decrease in general. There can be considerable challenges in confidently detecting long-term trends using historical reanalysis data, such as suggested in previous studies (Taszarek et al. 2021a; Pilguj et al. 2022), including noting the degree of interannual variability (e.g., from Fig. 1) as well as various other factors that might contribute to uncertainties in detecting trends such as potential changes over time in the available observations and their quality used to produce the reanalysis data. On the other hand, GCM simulations can feature considerable uncertainties as shown in previous studies including associated with their relatively coarse resolution (Kent et al. 2015; Huang et al. 2018; Taszarek et al. 2021a; Allan et al. 2023), while noting some benefits such as being spatiotemporally homogenous and enabling larger ensemble analyses over long time periods to help increase the signal to noise for trend detection. Consequently, it can sometimes be useful to consider contrasting data sources such as these to help provide complementary perspectives (as done in the following section using GCMs).

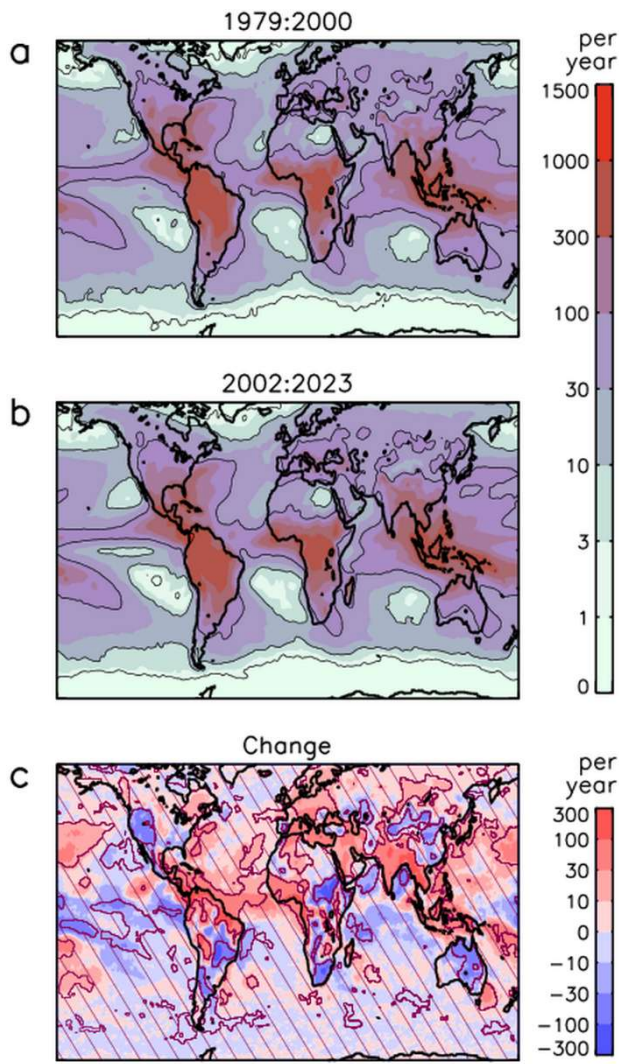

**Fig 6** The annual number of 6-hourly thunderstorm environments per grid cell is shown on average for the first half of the study period 1979 to 2000 (**a**) and the second half of the study period 2002 to 2023 (**b**) (noting the central year of 2001 is not used). The change from the earlier period (1979 to 2000) to the later period (2002 to 2023) is also shown (**c**), with hatching lines for locations where the change is not statistically significant (90% confidence level). Contour lines are shown for values of 1, 10, 100 and 1000 counts, with smoothing applied to the results shown in each panel (using a boxcar moving average of ±11 grid cells in latitude and longitude)

## 3.4 Climate model results

Results based on GCMs for the historical time period (Fig. 7a) show good consistency with those based on the ERA reanalysis (Fig. 6a). The results for the future projected time period suggest potential for increases in most locations (Fig. 7b,c). To explore this further, Table 4 shows the average values over the study region for the two components of the thunderstorm diagnostic (CAPE and WS06), indicating that the projected increase in thunderstorms indicated by the environment diagnostic is likely driven by increased CAPE. High values of CAPE are typically characterized by warm and moist air at low levels and/or air temperature decreasing rapidly with height, so for some general insight on the changes in CAPE, Table 4 also shows 850 hPa dewpoint (as one indicator of low-level moisture content) and temperatures at 850hPa and 500 hPa pressure levels (for general insight on vertical temperature gradients). The results in Table 1 show substantial increases in moisture, with a small decrease in vertical temperature gradient, indicative of vertical temperature gradients becoming more often associated with moist-adiabatic rather than dry-adiabatic gradients as more heat is released during condensation in convective updrafts. Spatial details of these projections are also shown in Appendix Figure 14 for the vertical temperature gradient and in Appendix Figure 15 for dewpoint, as well as Appendix Figure 16 for vertical wind shear which shows similar spatial patterns for the sign of change to those indicated based on ERA5 reanalysis (Appendix Figure 17).

Methods to detect climate change signals as distinct from other forms of variability can be based on when the signal grows larger than the standard deviation (Hawkins and Sutton 2012). The standard deviation of the thunderstorm environment occurrence frequency indicated by the diagnostic, based on annual totals at each grid cell, is about 28 on average throughout the region (from Fig. 1). This interannual variability is similar in magnitude to the change in thunderstorms per degree of global warming based on the diagnostic projections (an average increase of about 30 per year from Fig. 6), indicating that global warming above about 1°C might be needed for detecting trends in many regions. This suggests that detecting significant trends in Fig. 6 from the period 1979 to 2000 to the period 2002 to 2023 (i.e., 23 years later than the earlier period) could be difficult as the average global warming between those periods is less than 1°C (e.g., about 0.46°C based on data from https://data.giss.nasa.gov/gistemp/ accessed October 2025).

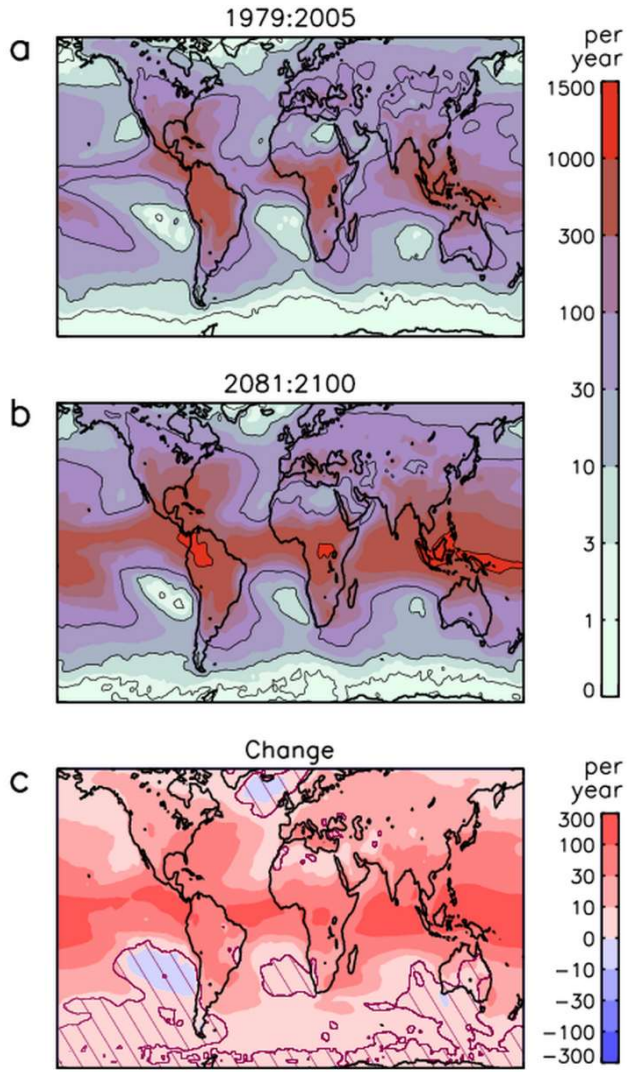

**Fig 7** The annual number of 6-hourly thunderstorm environments per grid cell is shown on average for the historical period (**a**) and future period (**b**) using the 12-member GCM ensemble. The change per degree of global warming is shown (**c**), with hatching lines for locations where the change is not statistically significant (90% confidence level). Contour lines are shown for values of 1, 10, 100 and 1000 counts, with smoothing applied to the results shown in each panel (using a boxcar moving average of ±11 grid cells in latitude and longitude)

**Table 4** GCM data averaged over the study region for CAPE, WS06, dewpoint at 850 hPa and temperatures at 850 hPa and 500 hPa. This is presented for historical and future time periods as well as the change per degree of global warming

|  | Historical | Future | Change per degree warming |
|---|---|---|---|
| CAPE | 434 J kg$^{-1}$ | 607 J kg$^{-1}$ | 45 J kg$^{-1}$ |
| Wind shear (WS06) | 13.9 m s$^{-1}$ | 13.6 m s$^{-1}$ | -0.1 m s$^{-1}$ |
| Temperature at 850 hPa | 7.1°C | 10.9°C | 1.0°C |
| Temperature at 500 hPa | -15.7°C | -11.4°C | 1.2°C |
| Dewpoint at 850 hPa | -0.7°C | 3.0°C | 1.0°C |

# 4    Discussion and summary

This study examined spatiotemporal variability in environments conducive to thunderstorm occurrence based on a dataset derived from reanalysis data from 1979 to 2023. The interannual variability was mapped throughout tropical and temperate regions of the world, showing that although the magnitude of the variability largely followed the magnitude of the mean values, the variability as a fraction of the mean occurrence frequency (i.e., the coefficient for variability) was larger in locations with smaller mean values.

ENSO was found to be the dominant climate mode influencing the variability of thunderstorm environments, similar to various previous studies (Allen and Karoly 2014; Pinto 2015; Dowdy 2016; Tippett and Lepore 2021; Malloy et al. 2023). The results presented here showed that this was particularly apparent for tropical regions and extratropical ocean regions, rather than at higher latitudes and in extratropical land regions. It was shown that 25% of the study region has a statistically significant relationship (at a 90% confidence level) between ENSO and thunderstorm environments, which is a more widespread region than occurs for the other modes of climate variability examined here (i.e., IOD, AO and AAO). Tropical ocean regions showed strong relationships with ENSO, particularly in the western and southwestern Pacific as shown in Fig. 3 where negative correlations occur (i.e., more thunderstorm environments on average for La Niña conditions associated with negative NINO3.4 anomalies than for El Niño conditions associated with positive NINO3.4 anomalies), consistent with the South Pacific Convergence Zone (SPCZ) extending further west on average for La Niña than El Niño conditions (Vincent et al. 2011; Dowdy 2020; Brown et al. 2020).

The IOD has a strong and widespread influence during the SON season, while noting this is closely related to the ENSO variability during this season (such as detailed in Tables 2 and 3). It was suggested that ENSO-IOD relationships could relate to feedback processes between the tropical Indian Ocean and Pacific Ocean linked through convection in the Maritime Continent region. Examining that possibility further provides scope for potential future research. The AO and AAO were found to influence some higher latitude locations (from Figures 4 and 5) where ENSO didn't have a strong influence (from Fig. 2), suggesting useful information can be provided by those polar annular modes of variability in addition to the coupled tropical modes of ocean-atmosphere variability such as ENSO and IOD.

The potential for long-term changes was examined in the number of thunderstorms environments, with results suggesting some regions with increases (15% of grid cell locations), as well as some regions with decreases (8% of grid cell locations) over the period 1979 to 2023. The regions of increase were mostly in the Northern Hemisphere, as well as some localized midlatitude regions in the western extents of the ocean basins in both hemispheres such as near southeast Australia. A hypothesis proposed here for those ocean regions of increase relates to strengthening ocean gyres producing greater sea surface temperature warming in western than eastern boundary currents (Cheng et al. 2022; Allan et al. 2023), which could result in enhanced source regions for warm moist air conducive to convection. However, future potential research could help explore this further, noting a range of uncertainties around detecting long-term trends as discussed through this study.

Information based on modelling was also considered, relating to potential changes in the number of thunderstorm environments in a warming world. This suggested more favorable conditions for thunderstorms might be likely in many regions, associated with more water vapor causing increased CAPE as the dominant factor over other changes such as a more stable vertical temperature gradient. This is consistent with theoretical expectations for potential increases in thunderstorm activity in a warmer world, including associated with Clausius Clapeyron considerations for increased water vapour concentrations (Romps et al. 2014; Hoogewind et al. 2017; Allen 2018; Dowdy 2020; Lepore et al. 2021; Cheng et al. 2022; Allan et al. 2023). However, if only considering reanalysis data the magnitude of interannual variability (from Fig. 1) is likely to be larger than the estimated magnitude of change over recent time periods such as 1979 to 2023 (from Fig. 6), making it hard to confidently detect long-term changes in favourable thunderstorm environments based on the period of historical reanalysis data used in this study.

The results presented here based on this diagnostic method suggest that some of the interannual variability in thunderstorm occurrence can be accounted for by various large-scale modes of atmospheric and ocean variability (including ENSO and others), with a small component potentially associated with long-term trends. However, a considerable portion of the interannual variability of thunderstorm environments (such as shown in Fig. 1) appears to be due to other causes that are not readily apparent, plausibly due in part to randomness given the inherent chaotic nature of the Earth system (Lorenz 1969). To examine the variability of the thunderstorm climatology in additional ways, further research based on this type of study methodology could also potentially consider using other modelling approaches and datasets, including applying the BTE diagnostic using finer-resolution downscaling data based on regional climate model output (Hoogewind et al. 2017; Brown et al. 2024). Further research could also potentially build on this study to examine individual convective hazard types (e.g., lightning, hail, extreme wind and rain/flood).

There can be various uncertainties based on using a single method for analysing the climatological variability of thunderstorm occurrence through the world, such as discussed in this study. It can therefore be useful to consider a broad range of lines of evidence from many studies, including as done in review studies such as Allen (2018). As such, the study results presented here are intended to be complementary to those of other studies, helping contribute as part of a broader range of information available on thunderstorms and climate variability throughout the world. Given the range of hazards that thunderstorms can cause, improved understanding of their climatological variability is expected to be useful in informing effective planning and decision making, including in relation to disaster risk reduction.

**Funding**

This research was supported through the Australian Centre of Excellence for 21$^{st}$ Century Weather, the Melbourne Energy Institute (MEI) and the Australian National Computational Infrastructure (NCI). Contribution of M. Taszarek was supported through a grant of the Polish National Science Centre (2020/39/D/ST10/00768).

**Competing Interests**

The authors have no relevant financial or non-financial interests to disclose.

**Data Availability**

Source data as used for this paper are available online (https://doi.org/10.5281/zenodo.16892383) as well as on request from the authors. Code used for data processing is publicly available here https://github.com/andrewbrown31/SCW-analysis/tree/master.

**Author Contributions**

This study was conceptualized by Andrew Dowdy, with data processing by Andrew Dowdy and Andrew Brown. The first draft of the manuscript was written by Andrew Dowdy as well as making all figures, with all authors contributing to subsequent drafts of the manuscript. All authors read and approved the final manuscript.

# Appendix

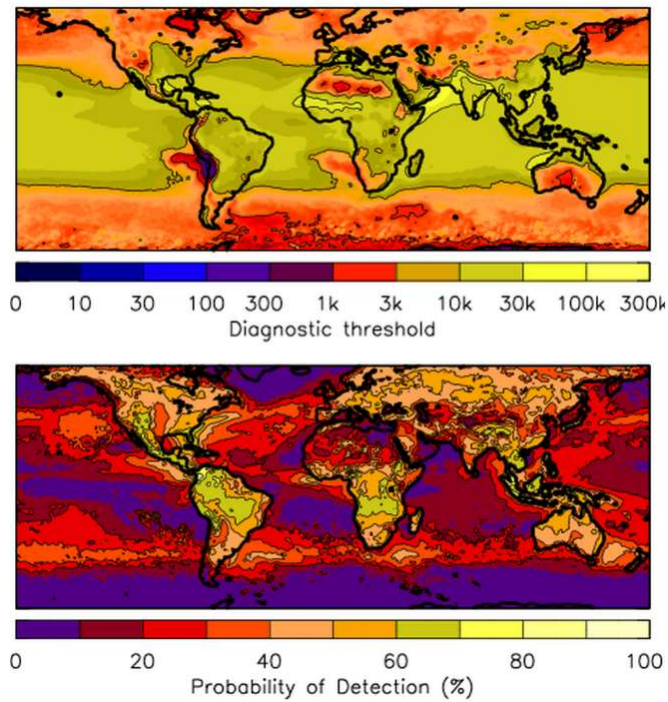

**Appendix Fig 8** Diagnostic threshold for BTE (in J m kg$^{-1}$ s$^{-1}$; upper panel) and the Probability of Detection, POD (in %; lower panel). We note that this figure is based on previous work such as published in Dowdy and Brown (2023)

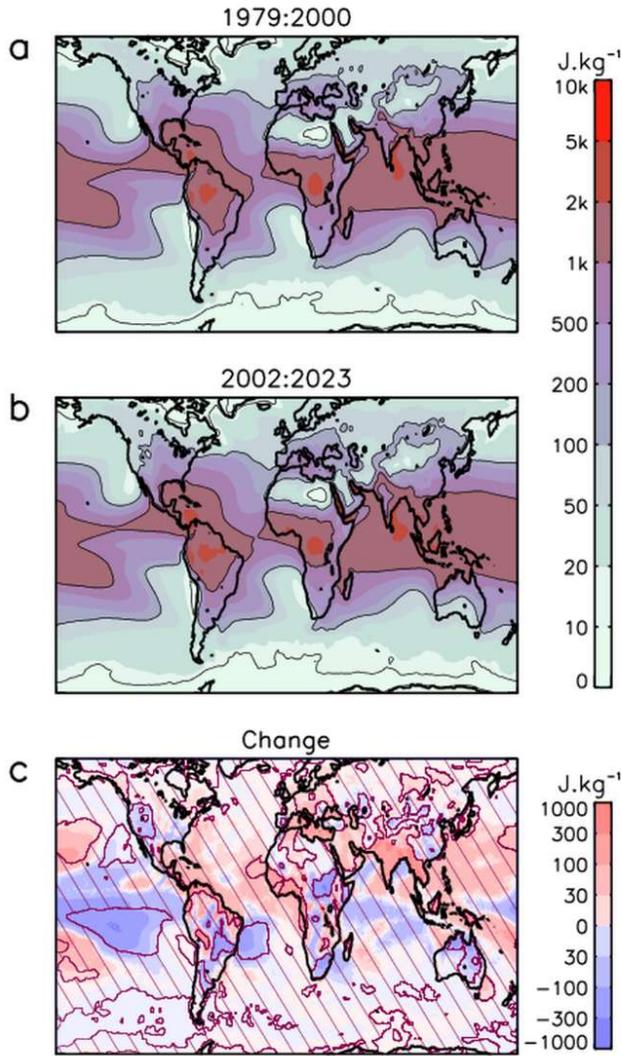

**Appendix Fig 9** CAPE is shown on average for the first half of the study period 1979 to 2000 (**a**) and the second half of the study period 2002 to 2023 (**b**) (noting the central year of 2001 is not used). The change from the earlier period (1979 to 2000) to the later period (2002 to 2023) is also shown (**c**), with hatching lines for locations where the change is not statistically significant (90% confidence level)

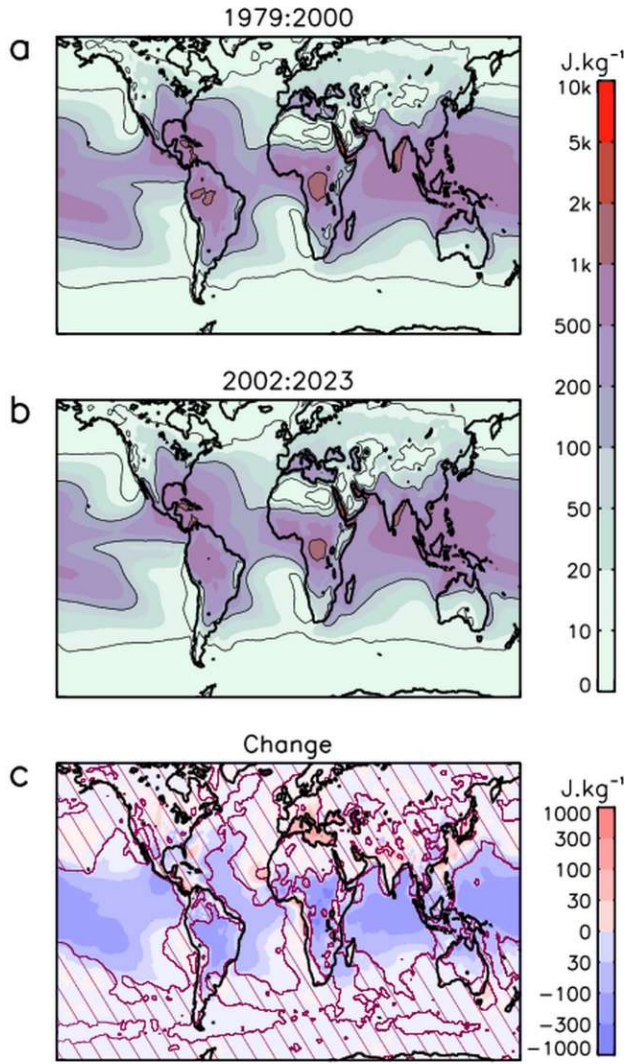

**Appendix Fig 10** As for Appendix Fig. 9, but using CAPE data obtained directly from the ERA5 reanalysis (referred to here as ERA5_CAPE), noting that these data are not used elsewhere in this study. ERA5_CAPE is shown on average for the first half of the study period 1979 to 2000 (**a**) and the second half of the study period 2002 to 2023 (**b**) (noting the central year of 2001 is not used). The change from the earlier period (1979 to 2000) to the later period (2002 to 2023) is also shown (**c**), with hatching lines for locations where the change is not statistically significant (90% confidence level)

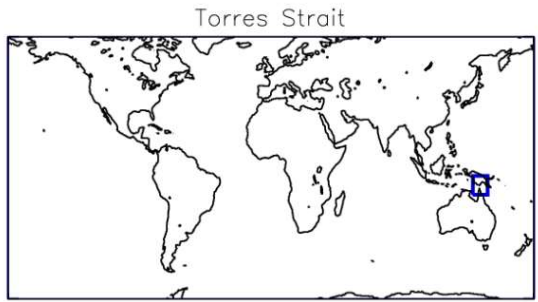

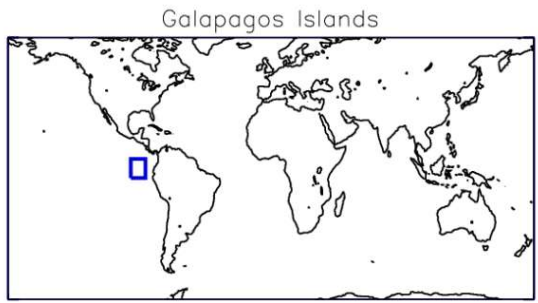

**Appendix Fig 11** Maps showing the location of the Torres Strait region (upper panel) referred to for Appendix Fig. 12, as well the Galagos Island region (lower panel) referred to for Appendix Fig. 13

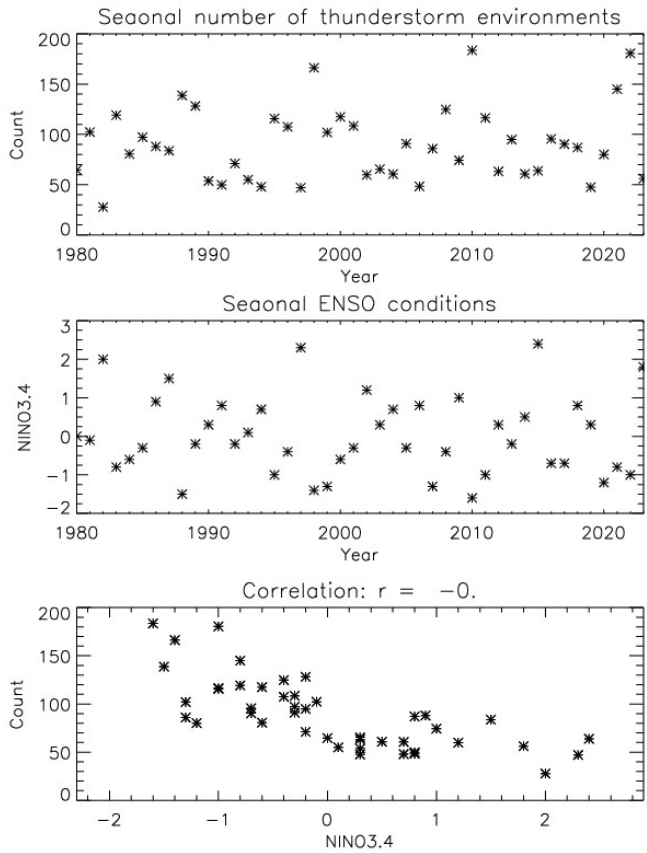

**Appendix Fig 12** Relationship between thunderstorm environments and ENSO for the Torres Strait region (as shown in Appendix Fig. 11) for the three-month period from September to November (referred to as the SON season). Time series are shown for the number of thunderstorm environments on average for a grid cell location in the Torres Stright region (upper panel), as well as for the average value of the NINO3.4 index (middle panel), in each SON season from 1980 to 2023. A scatterplot is also shown (lower panel) based on those seasonal time series shown here

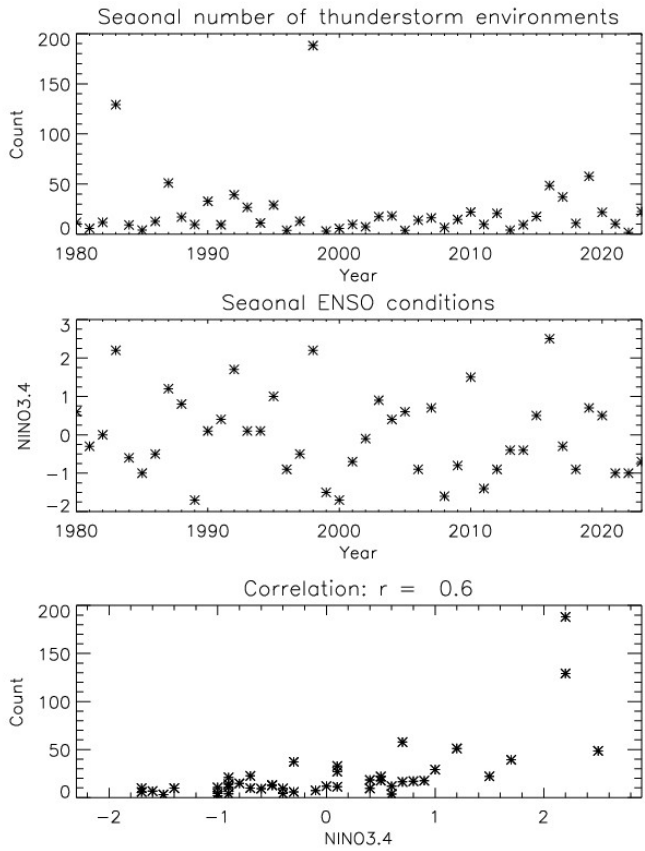

**Appendix Fig 13** Relationship between thunderstorm environments and ENSO for the Galapagos Island region (as shown in Appendix Fig. 11) for the three-month period from December to February (referred to as the DJF season, with December based on the previous year from January and February as detailed in Section 2). Time series are shown for the number of thunderstorm environments on average for a grid cell location in the Torres Stright region (upper panel), as well as for the average value of the NINO3.4 index (middle panel), in each SON season from 1980 to 2023. A scatterplot is also shown (lower panel) based on those seasonal time series shown here
27

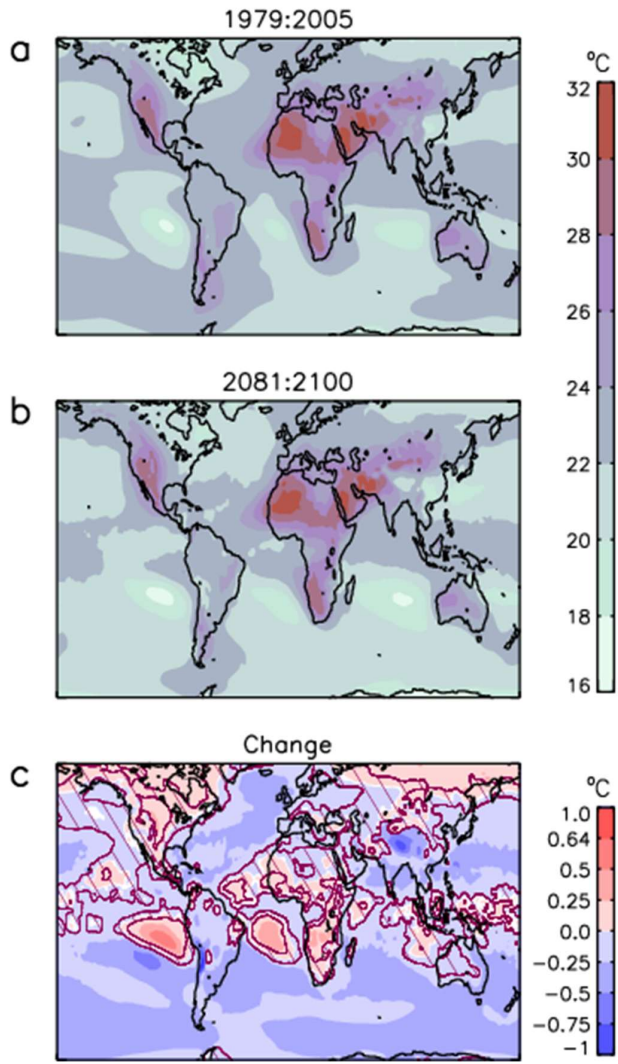

**Appendix Fig 14** Vertical temperature gradient from 850 hPa to 500 hPa, shown on average for the historical period 1979:2005 (**a**) and future period 2081:2100 (**b**) using the 12-member GCM ensemble. The change (i.e., future period – historical period) is shown (**c**), with hatching lines for locations where the change is not statistically significant (90% confidence level)

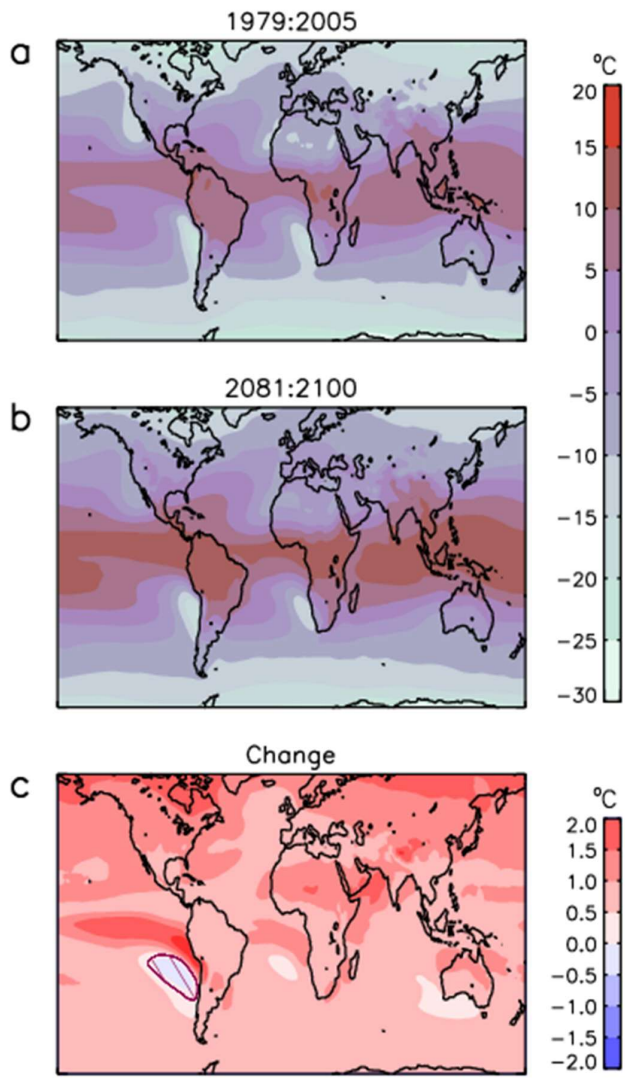

**Appendix Fig 15** Dewpoint temperature at 850 hPa shown on average for the historical period 1979:2005 (**a**) and future period 2081:2100 (**b**) using the 12-member GCM ensemble. The change (i.e., future period – historical period) is shown (**c**), with hatching lines for locations where the change is not statistically significant (90% confidence level)

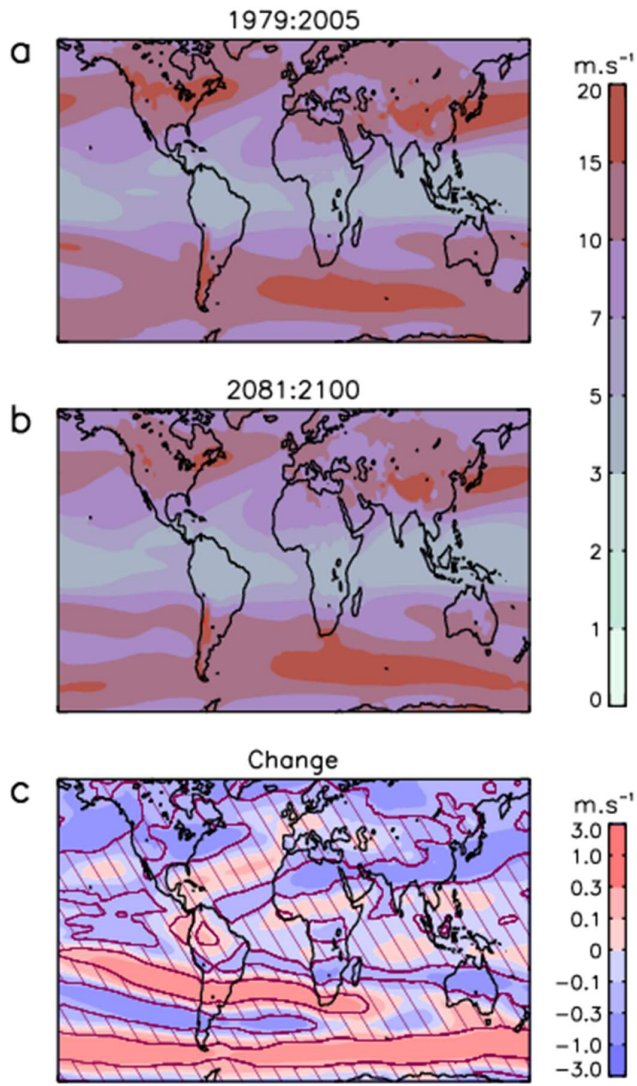

**Appendix Fig 16** Vertical wind shear (WS06) shown on average for the historical period 1979:2005 (**a**) and future period 2081:2100 (**b**) using the 12-member GCM ensemble. The change (i.e., future period – historical period) is shown (**c**), with hatching lines for locations where the change is not statistically significant (90% confidence level)

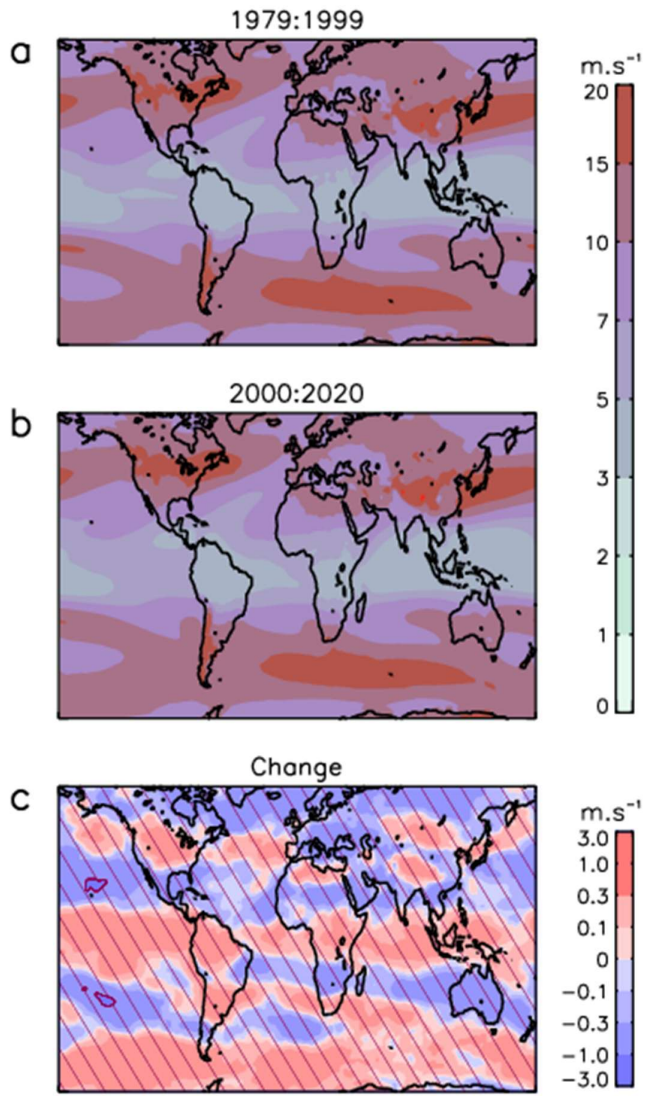

**Appendix Fig 17** Vertical wind shear (WS06) is shown on average for the first half of the study period 1979 to 2000 (**a**) and the second half of the study period 2002 to 2023 (**b**) (noting the central year of 2001 is not used). The change from the earlier period (1979 to 2000) to the later period (2002 to 2023) is also shown (**c**), with hatching lines for locations where the change is not statistically significant (90% confidence level)